# Wireless Power Transfer for High-precision Position Detection of Railroad Vehicles

Hyun-Gyu Ryu
Graduate School for Green Transportation
KAIST
Daejeon, Korea
hot6472@kaist.ac.kr

Dongsoo Har
Graduate School for Green Transportation
KAIST
Daejeon, Korea
dshar@kaist.ac.kr

*Abstract*—Detection of vehicle position is critical for successful operation of intelligent transportation system. In case of railroad transportation systems, position information of railroad vehicles can be detected by GPS, track circuits, and so on. In this paper, position detection based on tags onto sleepers of the track is investigated. Position information stored in the tags is read by a reader placed at the bottom of running railroad vehicle. Due to limited capacity of battery or its alternative in the tags, power required for transmission of position information to the reader is harvested by the tags from the power wirelessly transferred from the reader. Basic mechanism in wireless power transfer is magnetic induction and power transfer efficiency according to the relative location of the reader to a tag is discussed with simulation results. Since power transfer efficiency is significantly affected by the ferromagnetic material (steel) at the bottom of the railroad vehicle and the track, magnetic beam shaping by ferrite material is carried out. With the ferrite material for magnetic beam shaping, degradation of power transfer efficiency due to the steel is substantially reduced. Based on the experimental results, successful wireless power transfer to the tag coil is possible when transmitted power from the reader coil is close to a few watts.

I. INTRODUCTION

Position detection of railroad vehicles is critical for intelligent control of train operation. Efficient scheduling of individual trains is beneficial for avoidance of accident and enhancing the value of resource planning. Since scheduling of individual trains fundamentally depends on position detection of them, high precision in position detection is very crucial. Considering current trend of increased driving speed of railroad vehicles, importance of high precision position detection will be escalated. To achieve high precision in position detection, various approaches have been attempted so far.

Use of track circuits is one of the most common methods for railroad position detection. However, they require insulated rail joints for successful detection [1]. Eddy current sensor [2] supplies non-contact speed and distance measurement of railroad vehicles by detecting the magnetic variation along the track. The sensor works well at high speeds, but suffers difficulties when the vehicles accelerate or decelerate. Trip switch mounted on the overhead lines responds when a train with pantograph passes [3]. However, it cannot be used with other type of railroad systems operated by ground level or sub-ground level power lines. Axel counter [4] is also popularly adopted for position detection of railroad vehicles. It counts the number of wheel pairs passing the detection point. This causes a problem when its memory of the axle number is lost. GPS-based position detection is discussed in [5]. However, GPS signal is not detectable when railroad vehicles are in tunnels. In [6], position detection of running train is performed by the combination of GPS data, track curvature, and tachometer for secure position detection regardless of vehicle environment.

This paper presents a position detection method for high speed train by means of wireless power transfer to the tags installed on the sleepers of track. Reader coil is placed at the bottom of running railroad vehicles and tag coils transpond to the request of the reader coil. Position information of the tag is stored in the memory of it and the power necessary for the transmission of the position information to the reader coil is harvested by the tag coil. This paper is organized as follows. Section II gives the system model for wireless power transfer and Section III provides experimental results as well as emulation results. Results in Section III are obtained with reader coil and tag coil, each designed by impedance matching. Section IV concludes this work.

II. SYSTEM MODEL FOR WIRELESS POWER TRANSFER

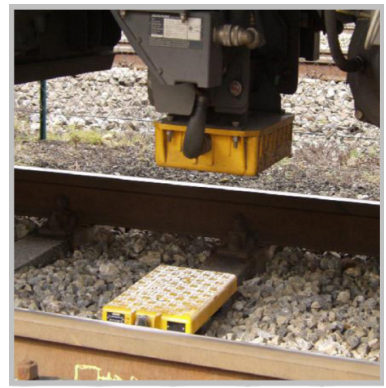

(a)

This research was supported by Ministry of Land, Infrastructure and Transport (MOLT) as Railroad Specialized Graduate School and a grant from Railroad Technology Research Program

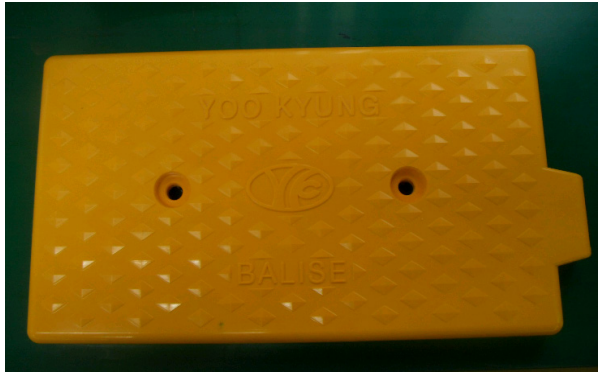

(b)

Figure 1. (a) Reader at railroad vehicle and (b) RFID tag on a sleeper of track for position detection (courtesy of Korea Railroad Research Institute)

Fig. 1 presents the reader placed at the bottom of railroad vehicle and the tag on a sleeper of the track. The reader transfers power wirelessly to provide energy for the tag. The energy harvested by the tag is used for transmission of position information stored in the memory of it. This procedure is carried out with the high train speed. Fig. 2 depicts the situation when the procedure for wireless power transfer and position information transmission is conducted. Before the reader is right above the tag, i.e., Δy takes negative values, the reader starts power transferring and the tag is initiating data transmission and continues to transmit until it is complete.

Carrier frequency for wireless power transfer is set to 27MHz. Fig. 3 shows the reader coil and tag coil used for this work. Reader coil for wireless power transfer is the inner loop of the two loops in Fig. 3(a). Position information obtained by the foregoing procedure is transmitted to the control center by a railroad communication system such as LTE-R OFDM system [7].

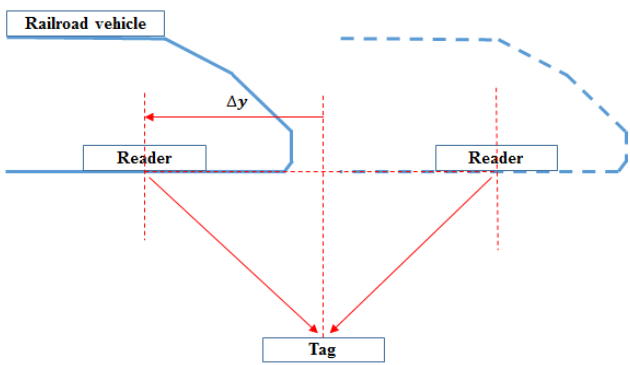

Figure 2. Situation for wireless power transfer and position information transmission.

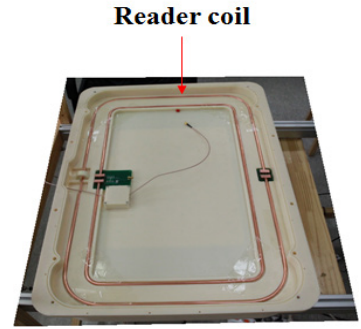

(a)

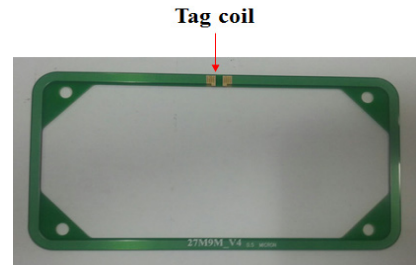

(b)

Figure 3. Reader coil and tag coil for wireless power transfer.

### III. EXPERIMENTAL RESULTS

Simulation by HFSS tool and actual measurements to evaluate the impact of ferromagnetic material (steel) modeling the track are performed for this work. Results of experiments are also effective with the presence of the steel at the bottom of railroad vehicles. Fig. 4 illustrates the configuration of the reader coil and the tag coil. Coil separation d is set to 0.5m to be close to actual separation between the track and the bottom of train. Location of the reader coil in Fig. 4 corresponds to Δy=0(see Fig. 2). When Δy is changed from 0 to 0.4m, received power at the tag coil varies as shown in Fig. 5. Plot in Fig. 5 is obtained by HFSS tool. Fig. 5 suggests that beyond Δy=0.4m received power will be less than 15%. This level of power transfer efficiency might be admissible with transferred power greater than a few watts and small power consumption by the tag during position information transmission.

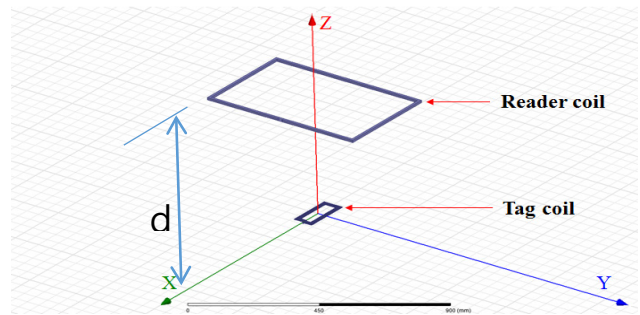

Figure 4. Setup of reader coil and tag coil for simulation

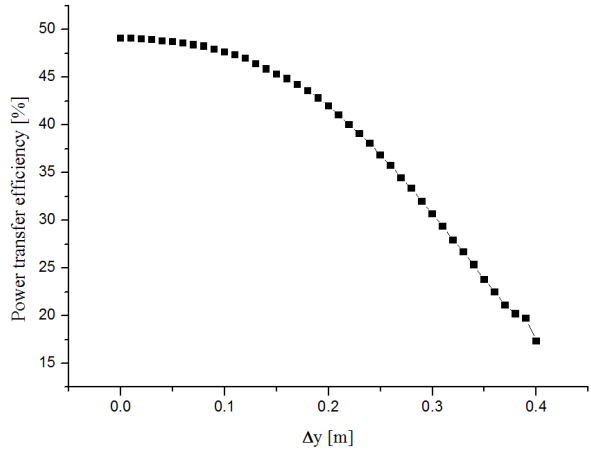

Figure 5. Variation of power transfer efficiency according to Δy.

Actual measurements with a network analyzer as a signal generator are executed as in Fig. 6. One port of the network analyzer is used as the transmitter (reader coil) and the other port is used as the receiver (tag coil). Power transfer efficiency is measured with and without steel. Also, power transfer efficiency with and without ferrite material (sheet) is measured. Ferrite material is known to be effective in magnetic beam shaping [8] to increase power transfer efficiency. Ferrite sheet is attached to the reader coil and the tag coil as shown in Fig. 7. The steel is placed behind the tag coil with a gap 0.05m or 0.1m. In Table I and Table II, Steel (X) and Ferrite(X), for example, indicate that no steel is placed behind the tag coil and no ferrite sheet is attached to reader coil and tag coil. As seen in Table I and Table II, impact of steel is substantially adverse to power transfer efficiency. However, with ferrite sheet attached to reader coil and tag coil, the power transfer efficiency becomes 48.08% from 31.84%, when the gap is 0.05m. In case of the gap=0.1m, it is 52% with ferrite sheet attached, which is significantly increased from 45.28% without ferrite sheet. From the results in Table I and Table II, it is judged that ferrite sheet is beneficial for sustaining power transfer efficiency with the presence of steel. It is noted that simulation results obtained by HFSS tool and experimental results are in good agreement.

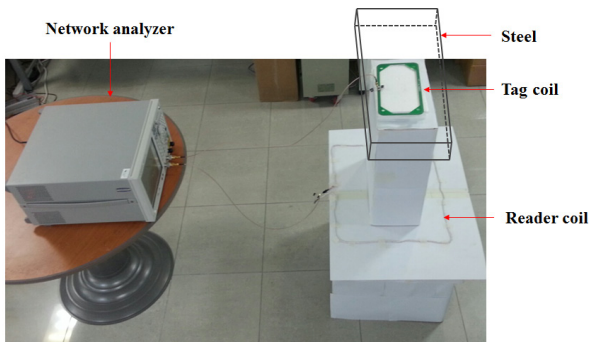

Figure 6. Experimental configuration of wireless power transfer.

TABLE I. POWER TRANSFER EFFICIENCY WHEN FERRITE SHEET AND STEEL ARE TAKEN INTO ACCOUNT. SEPARATION =0.5M AND GAP BETWEEN TAG COIL AND STEEL IS 0.05M.

|  | Ferrite (X) | Ferrite (O) |
|---|---|---|
| Steel (X) | 53.57% | 52.48% |
| Steel (O) | 31.84% | 48.08% |

TABLE II. POWER TRANSFER EFFICIENCY WHEN FERRITE SHEET AND STEEL ARE TAKEN INTO ACCOUNT. SEPARATION =0.5M AND GAP BETWEEN TAG COIL AND STEEL IS 0.1M.

|  | Ferrite (X) | Ferrite (O) |
|---|---|---|
| Steel (X) | 53.57% | 52.48% |
| Steel (O) | 45.28% | 52.00% |

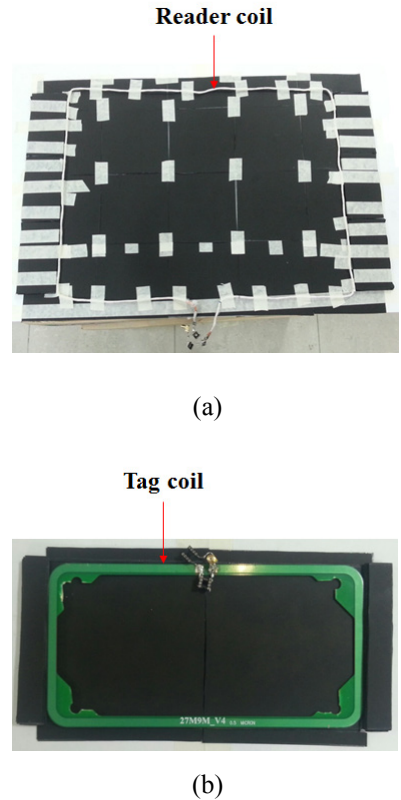

Figure 7. (a) Reader coil with ferrite sheet and (b) tag coil with ferrite sheet.

IV. CONCLUSION

For successful operation of high speed train, high precision position detection is critical for intelligent control of train operation. This paper presents wireless power transfer to get the position information stored in the tags on the sleepers of the track. Basic mechanism in wireless power transfer is magnetic induction and power transfer efficiency is discussed with simulation results and experimental results. Since power transfer efficiency is significantly affected by the steel, magnetic beam shaping by ferrite material is carried out. It is shown here that with the ferrite material for magnetic beam shaping, degradation of power transfer efficiency due to the steel is substantially reduced. Based on the results, successful wireless power transfer to the tag coil is possible when transmitted power from the reader coil is close to a few watts and power consumption of the tag for position information transformation is small.